\documentclass[prl,a4paper,twocolumn,footinbib,showpacs]{revtex4}

\usepackage[ansinew]{inputenc}
\usepackage{amssymb}
\usepackage{amsmath}
\usepackage{amsbsy}
\usepackage{graphicx}

\begin{document}
\title{Kinetic step bunching instability during surface growth}

\author{Thomas Frisch}\email{frisch@irphe.univ-mrs.fr}
   \affiliation{%
   Institut de Recherche sur les Phénomènes Hors \'Equilibre,
   UMR 6594, CNRS, Université de Provence, Marseille, France}

   \author{Aberto Verga}\email{verga@irphe.univ-mrs.fr}
   \affiliation{%
   Institut de Recherche sur les Phénomènes Hors \'Equilibre,
   UMR 6594, CNRS, Université de Provence, Marseille, France}

\date{\today}

\begin{abstract}
We study the step bunching kinetic instability in a growing crystal
surface characterized by anisotropic diffusion. The instability is
due to the interplay between the elastic interactions and the
alternation of step parameters. This instability is predicted to
occur on a vicinal semiconductor surface Si(001) or Ge(001) during
epitaxial growth. The maximal growth rate of the step bunching
increases like $F^{4}$, where $F$ is the deposition flux. Our
results are complemented with numerical simulations which reveals a
coarsening behavior on the long time for the nonlinear step
dynamics.
\end{abstract}

% 05.70.Ln, 47.20.Hw, 68.35.Bs, 68.35.Rh, 68.55.Ac, 61.16.Ch, 68.65.–k

\maketitle

The field of surface growth on semiconductors is very active due to
its importance for both technological applications and fundamental
science \cite{stangl04,pimpinelli98,saito98}. Under non-equilibrium
growth a variety of experiments reveal rich crystal morphologies
resulting from the nonlinear evolution of step flow instabilities
\cite{jeong99,yagi01}. The kinetic and elastic effects drive the
system towards self-organized states, characterized by the
appearance of ordered structures on the vicinal surfaces. This
self-organization can be exploited in semiconductor nanotechnology,
for the development of devices having interesting quantum properties
\cite{shchukin99,brunner02}. A basic mechanism for pattern formation
in vicinal semiconductor surfaces is the step bunching instability
\cite{tersoff95,pierre-louis03,krug05} whose origin is commonly
attributed to the presence of impurities, to the inverse
Ehrlich-Schwoebel effect, or to electro-migration (see
\cite{krug04,pierre-louis04} and references therein). This Letter is
motivated by the molecular beam epitaxy (MBE) experiments on Si(001)
in which it was observed a new type of kinetic instability leading
to the formation of step bunches
\cite{schelling99,myslivecek02a,pascale03}. Microscopic kinetic
Monte-Carlo simulation showed that in the case of Si(001), step
bunching is due to the coupling between diffusion anisotropy and
differences in step kinetic parameters \cite{myslivecek02b}. Here we
provide a macroscopic instability mechanism for the step bunching
instability that does not require any inverse Ehrlich-Schwoebel
effect. We show that the interplay between the elastic step
interactions and the alternation of kinetic parameters,
characteristic of the Si(001) vicinal surface, induces a finite
wavelength instability with maximal growth rate increasing as
$F^{4}$  ($F$ is the deposition flux). Our results are complemented
by numerical simulations which reveal a coarsening behavior on the
long time for the non-linear step dynamics.

\begin{figure}
\centering
\includegraphics[width=0.48\textwidth]{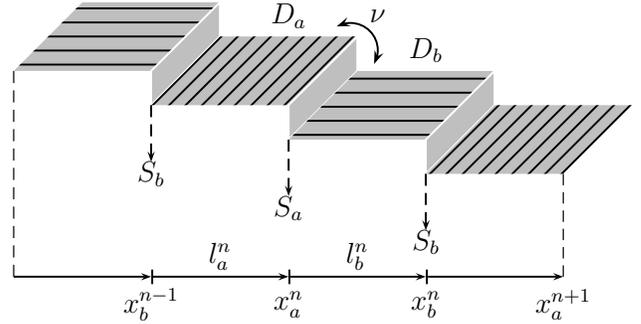}
\caption{Sketch of the Si(001) vicinal surface showing the
    alternation of terraces and steps $S_a$ and $S_b$. $D_a$ and $D_b$
    are the surface diffusion coefficients, and $\nu$ the upper and
    lower terrace kinetic attachment coefficient. The horizontal axis
    shows terrace lengths $l^n_{a}$,\, $l^n_{b}$ and step positions $x^n_{a}$,\, $x^n_{b}$.}
\label{fig1}
\end{figure}

The Si(001) vicinal surface consists of a periodic sequence of
terraces where rows of $2\times1$ dimerised adatoms (terrace of type
$a$) alternate with $1\times2$ dimerised adatoms  (terrace of type
$b$), see Fig.~\ref{fig1}. On the reconstructed surface adatoms
diffuse preferentially along dimer rows, giving rise to an
anisotropic diffusion. Therefore, the steps separating the terraces
are of two kinds. The $S_a$ step is rather straight while the $S_b$
step is very corrugated \cite{zandvielt00}. The significant
difference in the attachment kinetics between smooth and rough steps
allows us to choose the kinetic coefficient $\nu_a=\nu$ of the $S_a$
step to be finite and the one of the $S_b$ step to be infinite
($\nu_b\rightarrow\infty$) \cite{sato03a}.

\begin{figure}
\centering
\includegraphics[width=0.48\textwidth]{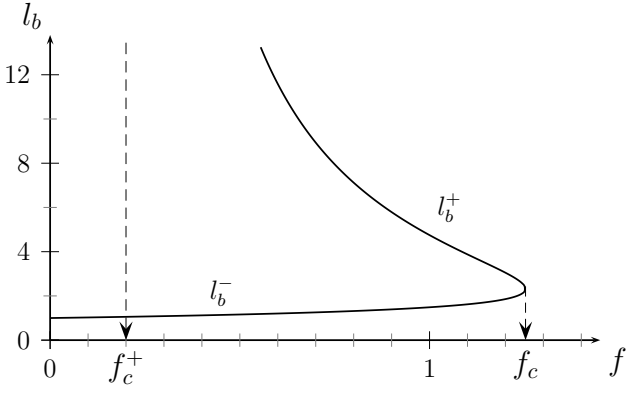}
\caption{Terrace size $l_b$ as a function of the flux $f$
    (nondimensional) for the fixed point solution with $\alpha_a=1$ and
    $\alpha_b=0.1$. For $f>f_{c}=4(2+\alpha_a+\alpha_b)
        (3+2\alpha_a-2\sqrt{(1+\alpha_a)(2+\alpha_a)}\,)$
    no stationary solution exists. For
    $f<f_{c}^+=4\alpha_b/(1+\alpha_a+\alpha_b)$
    the upper branch disappears.}
\label{fig2}
\end{figure}

Let us denote by $x_a^n(t)$ and $x_b^n(t)$ the positions at time $t$
of steps $S_a$ and $S_b$ respectively, the terrace sizes are
$l_a^n(t)=x_a^n(t)-x_b^{n-1}(t)$ and
$l_b^n(t)=x_b^{n}(t)-x_a^{n}(t)$, and $v_a^n(t)=\dot x_a^n(t)$,
$v_b^n(t)=\dot x_b^n(t)$ are the step velocities, where
$n=1,2,\ldots$ numbers the terrace (Fig.~\ref{fig1}). During the
growth, the adatom concentration on each terrace obeys the following
diffusion equations \cite{burton51}.
\begin{equation}\label{bcf}
    D_a\frac{\partial^2}{\partial x^2} C_a^n(x,t)=-F\,,
        \; \mathrm{and} \;
    D_b\frac{\partial^2}{\partial x^2} C_b^n(x,t)=-F\,,
\end{equation}
where $C_a^n(x,t)$ and $C_b^n(x,t)$ are the adatom concentration on
each terrace, $D_a$ and $D_b$ the diffusion coefficients, and $F$
the deposition flux. We assume that the desorption of adatoms is
negligible, since step bunching is observed at temperatures less
then $1200\,$K; we also neglect transparency of steps. Equations
(\ref{bcf}) are subject to boundary conditions at each step:
\begin{eqnarray}
    C_a^n(x_b^{n-1}) &=& C_{eq,b}^n            \label{bc1}  \\
    D_a\frac{\partial C_a^n}{\partial x}(x_a^n)&=&
        -\nu\left[C_a^n(x_a^n)-C_{eq,a}^n\right]   \label{bc2} \\
    D_b\frac{\partial C_b^n}{\partial x}(x_a^n)&=&
        \nu \left[C_b^n(x_a^n)-C_{eq,a}^n\right]       \label{bc3} \\
    C_b^n(x_b^n)&=&C_{eq,b}^{n+1}\,,  \label{bc4}
\end{eqnarray}
where $\nu$ is the step kinetic coefficient of step $a$;
$C_{eq,a}^n$ and $C_{eq,b}^n$ are the adatom equilibrium
concentration at each step. The velocity of each step is given by
the conditions,
\begin{eqnarray}
    v_a^n &=& \Omega D_b \frac{\partial C_b^n}{\partial x}(x_a^n)-
        \Omega D_a\frac{\partial C_a^n}{\partial x}(x_a^n)
      \label{vab1}  \\
    v_b^n &=& \Omega D_a \frac{\partial  C_a^{n+1}}{\partial x}(x_b^n)-
        \Omega D_b\frac{\partial C_b^n}{\partial x}(x_b^n)\,,
        \label{vab2}
\end{eqnarray}
with $\Omega$ the unit atomic surface. The adatom equilibrium
concentrations  are determined by the elastic interactions between
steps mediated by the terrace reconstruction
\cite{alerhand88,muller04,sato03a},
\begin{eqnarray}
    C_{eq,b}^n &=& C_0+E\left(\frac{1}{l_a^n}-
            \frac{1}{l_b^{n-1}}\right)\\
    C_{eq,a}^n &=& C_0+E\left(\frac{1}{l_b^n}-
            \frac{1}{l_a^{n}}\right)
\end{eqnarray}
where $E=\Omega C_0 A/k_BT$ ($C_0$ is the uniform equilibrium
concentration, $k_B$ the Boltzmann constant, $T$ the temperature,
and $A$ is an energy per unit length related to the stress and
elastic constants of the medium \cite{alerhand88}). The system
behavior is controlled by three independent nondimensional
parameters: $f=Fl_a^2/\nu E$, $\alpha_{a}=\nu l_a/D_{a}$ and
$\alpha_{b}=\nu l_a/D_{b}$. We set the unit length to be the initial
size of terrace $a$, $l_a=1$. With this convention the velocity is
measured in units of $\nu E\Omega/l_a$. Under normal experimental
conditions ($\Omega F<1\,\mathrm{s^{-1}}$,
$A\approx5\,10^{-12}\,\mathrm{J\,m^{-1}}$, $\Omega C_0\approx
10^{-2}$) $f\lesssim1$, $\alpha_a\gtrsim1$ and $\alpha_b\ll1$.

We first look for a uniform train of steps with $l_a^{n}=1$ and
$l_b^{n}=l_b$ for all $n$, traveling at a constant velocity. The
relative velocity is
\begin{equation}\label{dv}
    v_b-v_a=\frac{4 -l_b(4-f)}
     {l_b(1+\alpha_a)}+
    \frac{4-l_b(4-l_bf)}{l_b(1+l_b\alpha_b)}\,.
\end{equation}
It vanishes for terraces of size $l_b$ giving by
\begin{equation} \label{lb}
    l_b^{\pm}=2\frac{2-f/4+ \alpha_a - \alpha_b \pm
        \sqrt{\Delta}}{f(1+\alpha_a + \alpha_b-4\alpha_b/f)}\,,
\end{equation}
where $\Delta=(2 -f/4 +
\alpha_a-\alpha_b)^2-f(2+\alpha_a)(1+\alpha_a+\alpha_b-
4\alpha_b/f)$. Depending on the values of the parameters
$(\alpha_{a},\alpha_{b},f)$ different situations may arise. The two
solutions $l_b^{\pm}$ represent terraces of nearly equal size
($l_b^-$), or double steps ($l_b^+$) for which terraces of type $a$
almost disappear (Fig.~\ref{fig2}). This two branches exist
($\Delta>0,\,l_b>0$) for fluxes lower than $f_c$. The upper branch
characterizing the double step state, exists in the interval
$(f_c^+,f_c)$, as shown in Fig.~\ref{fig2}. This interval vanishes
for the values of $\alpha_b$ satisfying
$\alpha_b^2>(1+\alpha_a)(2+\alpha_a)$.

\begin{figure}
\includegraphics[width=0.48\textwidth]{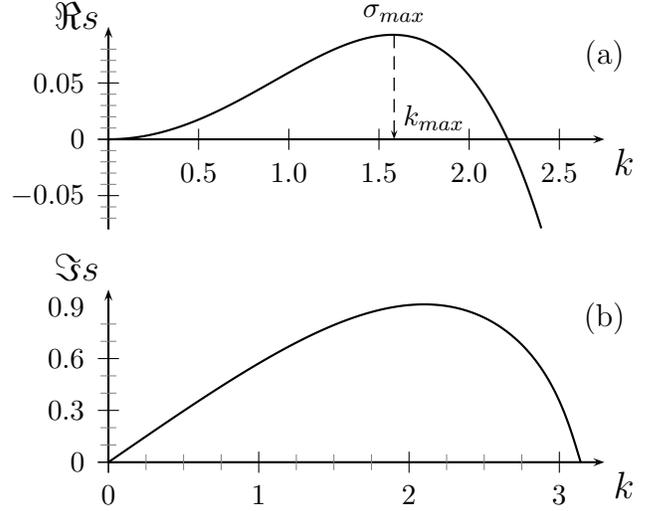}
\caption{The growth rate $s$ versus $k$, (a) real and (b) imaginary
    parts, for the $l_b^-$ branch with $\alpha_a=1$, $\alpha_b=0.1$, and
    $f=1.2<f_c$.}
\label{fig3}
\end{figure}

\begin{figure}
\includegraphics[width=0.48\textwidth]{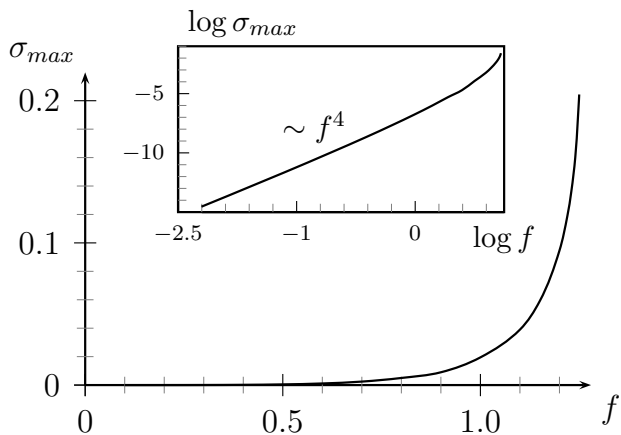}
\caption{Maximum growth rate $\Re s=\sigma_{max}$ versus  $f$, for
    the $l_b^-$ branch with $\alpha_a=1$, and $\alpha_b=0.1$. Inset in
    log-log scale showing the $f^4$ law for small $f$.}
\label{fig4}
\end{figure}

We investigate now the stability of a train of steps traveling at
the constant velocity $V=v_{a}(l_b)=v_{b}(l_b)$, where $l_b$ is one
of the solutions (\ref{lb}). The perturbed step positions are given
by $x_a^n=Vt+n(1+l_b)+X_{a}^n$ and $x_b^n=Vt+n(1+l_b)+l_b+X_{b}^n$.
We expand the step velocities up to first order in the perturbation
$X_{a,b}^n(t)=X_{a,b}\exp(st+\mathrm{i}nk)$, where $s$ is the
(complex) growth rate and $k$ the wavenumber. After some algebra, we
obtain the following expression for the growth rate
$s=\sigma+\mathrm{i}\omega$, in the limit of small $f$ and $k$, for
the lower branch $l_b^-$,
\begin{equation}\label{sigma}
    \sigma=Af^2k^2-Bk^4\,,\quad
    \omega=\frac{1}{2}fk-Cfk^3
\end{equation}
where
\begin{eqnarray*}
    A&=&\frac{1}{16}
        \frac{(\alpha_a+\alpha_{{b}}+3)}{
        (\alpha_a+\alpha_b+2)^2}    \\
    B&=&\frac{1}{1536(\alpha_a+\alpha_b+2)^{3}} \times\\
    && \left[768(1+\alpha_a+\alpha_b)+
        192(\alpha_a+\alpha_b)^2- \right. \\
     &&   \left.
        24(8+6\alpha_a+14\alpha_b+\alpha_a^2+
            6\alpha_a\alpha_b+5\alpha_b^2)f\right.-\\
    &&   \left.(72+86(\alpha_a+\alpha_b)+19\alpha_a^2+
            80\alpha_a\alpha_b-11\alpha_b^2)f^2\right]
             \\
    C&=& \frac{1}{384}
        \frac{8\alpha_a+8\alpha_b+16-9(\alpha_a-\alpha_b)f}{
        2+\alpha_a+\alpha_b}\,.
\end{eqnarray*}

Figure~\ref{fig3} shows the real and imaginary parts of $s$ for
typical values of the parameters. The imaginary part $\omega=\Im
s\sim fk$ is associated with the propagation of compression waves.
The upper branch $l_b^+$ displays a similar behavior, showing that
both uniform solutions are unstable. The growth rate $\sigma=\Re s$
of this instability increases like $f^2k^2$ for small $k$ and is
stabilized at larger $k$ by elastic effects (independent of $f$).
The most unstable mode $k_{max}$ increases linearly with the flux
$k_{max}\sim f$, and its corresponding growth rate scales like
$\sigma_{max}\sim f^4$, as can be seen from Eq.~(\ref{sigma}) and
illustrated in Fig.~\ref{fig4}. Therefore, for small $f$, the
instability is almost vanishing. We have checked that this result is
not qualitatively changed if a finite kinetic coefficient $\nu_b$
for the $b$ step is included into the model. In the particular case
$\nu_a=\nu_b$, the growth of the step flow instability vanishes,
independently of the $D_a/D_b$ ratio.

This step bunching instability is a consequence of kinetic and
elastic coupling under crystal growth conditions. The asymmetry of
kinetic coefficients, related to the alternation of smooth and rough
steps, introduces a supplementary elastic coupling between
\emph{non-neighboring} terraces. Their combined action breaks the
symmetry between the ascending and descending adatoms currents. In
contrast to the usual inverse Ehrlich-Schwoebel effect, the
unbalanced currents concern the interaction between two double
terraces.

\begin{figure}
\includegraphics[width=0.48\textwidth]{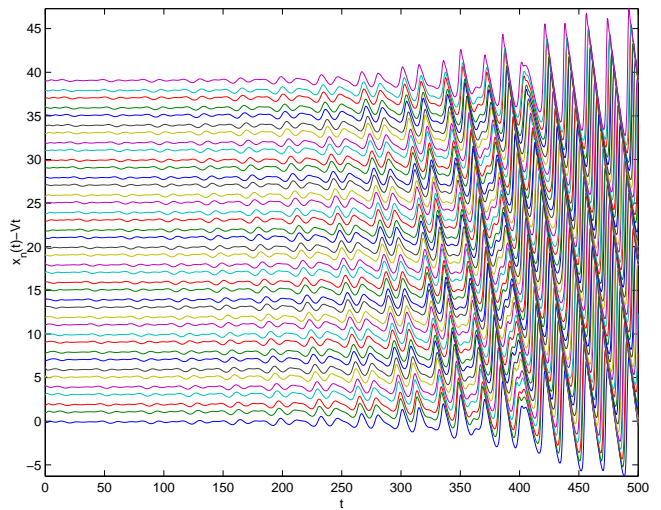}
\caption{Initial spacetime evolution of a train of $N=40$ steps
    $x_n(t)$ (mean velocity subtracted). Numerical resolution of
    Eqs.~(\protect\ref{vabn1}-\protect\ref{vabn2})
    with $\alpha_a=1$, $\alpha_b=0.1$, and $f=1.2$.}
\label{fig5}
\end{figure}

\begin{figure}
\includegraphics[width=0.48\textwidth]{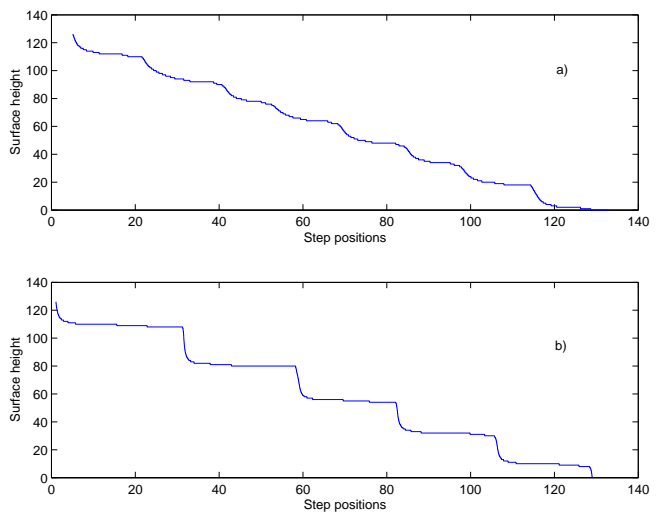}
\caption{Snapshots of the surface during step bunching for
    $\alpha_a=1$ and $\alpha_b=0.1$, $f=1.2$ and $N=128$. (a) The
    modulated step pattern at $t=400$; (b) the coarsening regime at
    $t=1000$.}
\label{fig6}
\end{figure}

We can obtain the non-linear evolution of the step positions by
numerical integration of the set of differential equations
(\ref{vab1}-\ref{vab2}). These equations can be written explicitly
using (\ref{bcf}) and its associated boundaries conditions
(\ref{bc1}-\ref{bc4}), they read:
\begin{eqnarray}
    v_a^n &=& \frac{\mu_a^n+\frac{f\alpha_b{l_b^n}^2}{2}}{1+\alpha_bl_b^n}-
        \frac{\mu_b^n+f l_a^n (2+ \frac{3\alpha_a l_a^n}{2} )}{1+\alpha_al_a^n}
            \label{vabn1}\\
    v_b^n &=&\frac{\mu_b^{n+1}+ fl_a^{n+1}(1+\frac{\alpha_a l_a^{n+1}}{2})}%
            {1+\alpha_a l_a^{n+1}}-
        \frac{\mu_a^n+fl_b^n(1+\frac{3\alpha_b l_b^n}{2})}%
            {1+\alpha_b l_b^n} \,,\nonumber \\
            &&\label{vabn2}
\end{eqnarray}
where $\mu_a^n=1/l_a^{n+1}+1/l_a^{n}-2/l_b^{n}$, and
$\mu_b^n=1/l_b^{n-1}+1/l_b^{n}-2/l_a^{n}$. The typical evolution of
the step flow on the vicinal surface is shown in Figs.~\ref{fig5}
and \ref{fig6}. We choose as initial conditions different sets of
slightly perturbed equal size terraces. The amplitude of the
perturbations increases during the evolution while compression waves
propagate as predicted by the linear analysis. The further evolution
of the system shows the formation of bunches of steps separated by
wider terraces. A coarsening regime sets in, characterized by the
coalescence of these step bunches into larger ones. Ultimately the
long time evolution leads to a surface composed of a single
macro-step. We have checked that these results are not affected by
the choice of initial conditions or by the size of the system. This
behavior is illustrated in Fig.~\ref{fig6} by numerical simulations
depicting the step train at two different times for $f<f_c$. We have
also checked that this scenario remains valid when $f>f_c$.
Figure~\ref{fig6}(a) shows a modulated step pattern characteristic
of the incipient formation of step bunches. Figure~\ref{fig6}(b)
shows the further evolution of these bunches during the coarsening
regime. A variety of models of step bunching shows that the mean
number of steps in a single bunch obeys a power law $\langle
n(t)\rangle\sim t^{\beta}$ \cite{tersoff95,krug04}. In the present
case this behavior is expected to be more complex due to the
presence of oscillations in the step size. This point deserves a
specific investigation using a continuous description of the step
dynamics, more appropriated for the study of the long time
evolution.

In conclusion, we have proposed a kinetic mechanism of step bunching
instability under non-equilibrium growth of a vicinal surface of
Si(001)-like crystals. The essential ingredients are the alternation
of rough and smooth steps, having different diffusion and attachment
coefficients, and the coupling of non-neighboring terraces by
elastic interactions. We have found that the rate of growth of the
instability scales like $f^4$. It would be interesting to
experimentally test this behavior, using for example measurements of
the surface roughness evolution. Some indirect evidence supporting
the present instability mechanism are provided by experiments on
Si/Ge growth \cite{schelling01}.  The addition of a small quantity
of Ge atoms produces a substantial modification of the steps
properties, and in particular shades off the distinction between
rough and smooth steps. As a consequence, a stable step flow regime
sets in even though the diffusion coefficient of successive terraces
are still different. A generalization of the present model  to two
dimensions is currently under progress. We expect a rich
phenomenology due to the coupling between a meandering instability
induced by the alternation of the step parameters \cite{sato03}, the
step bunching instability and the anisotropy of surface diffusion
\cite{danker04}.

\acknowledgements{ We would like to thank Isabelle Berbezier,
Antoine Ronda, Jean-Jacques Métois, Andrés Sa\'ul and Olivier
Pierre-Louis for stimulating discussions.}

\bibliography{mondocument}

\end{document}